\documentclass[12pt]{article}  

\def\ahalf{{\textstyle{1\over2}}}

\def\be{\begin{equation}}
\def\ee{\end{equation}}
\def\bea{\begin{eqnarray}}
\def\eea{\end{eqnarray}}

\def\ie{{\it i.e.}}

\def\v#1{{\bf#1}}

\def\'#1{{\accent19\ifx #1i \i\else #1\fi}}

\newcommand{\bfalpha}{\mbox{\boldmath$\alpha$\unboldmath}}

\newcommand{\bfgamma}{\mbox{\boldmath$\gamma$\unboldmath}}

\newcommand{\bfpi}{\mbox{\boldmath$\pi$\unboldmath}}
\newcommand{\bfrho}{\mbox{\boldmath$\rho$\unboldmath}}
\newcommand{\ocal}{\mathcal{O}}

\title{The many body problem in relativistic quantum mechanics}

\author{Marcos Moshinsky\thanks{Member of El Colegio Nacional and Sistema Nacional de Investigadores}
 and Anatoly Nikitin$^a$}

\date{\small Instituto de F\'{\i}sica\\
 Universidad Nacional Aut\'onoma de M\'exico\\
  Apartado Postal  20-364,
 01000 M\'exico D.F., M\'exico\\
 moshi@fisica.unam.mx\\
 $^a$ {Permanent Address: Institute of Mathematics \\ Ukranian Academy of
Sciences,\\ 3 Tereshchenkivska Str., 01601 Kyiv, Ukraine}}

\begin{document}

\maketitle

\begin{abstract}

We discusse a relativistic Hamiltonian for an n-body problem in
which all the masses are equal and all spins take value 1/2.
In the frame of reference in which the
total momentum $\v P=0$, the Foldy-Wouthuysen
transformation is applies and  the positive energy
part of the Hamiltonian is separated. The Hamiltonian with
unharmonic oscillator potential is applied to describe mass differences
for  charmonium and bottonium states.
\end{abstract}

\newpage

\section{Introduction}

In the development of non-relativistic quantum mechanics the
contribution of Schr\"odinger  was not only in writing a
single particle wave equation that, for a Coulomb potential, leads
to the spectrum of the hydrogen atom, but also that his formalism
could be extended immediately to a system of n-particles in
configuration space. Thus wave mechanics from the very beginning
was able to deal with systems of many particles in interaction as
happens for electron in atoms or nucleons in nuclei.

The non-relativistic many body Hamiltonian, in the absence of
spins, was composed of the sum of the kinetic energies
$\sum^n_{s=1}(p^2_s/2m_s)$ of the $n$ particles together with an
appropriate potential interaction between them. Many refined
mathematical methods \cite{2} were developed to obtain the
eigenvalues and eigenstates of these Hamiltonians.

The corresponding development did not take place in the theory of
many body problems in relativistic quantum mechanics. It is true
that almost simultaneously to the development of non-relativistic
quantum mechanics Dirac \cite{dirac2} introduced his famous
relativistic wave equation for a single particle. The extension of
this equation to many bodies found two main obstacles. One of them
was the need to formulate the wave equation in a form invariant
under the transformations of the Poincar\'e group, thus
guaranteing its relativistic character. In particular relativistic
invariance presupposes a certain symmetry between space and time
variables, which causes the appearance of multi-time variables in
an n-body theory and thus is an aspect that we will correct
without violating Poincar\'e invariance.

The other obstacle was that in relativity the relation between
energy $E$ and momentum $p$ of a particle of mass $m$ is
        \be
        E^2 = p^2c^2+m^2c^4
        \label{mk01}
        \ee
instead of the $E=p^2/2m$ of the non-relativistic case.

Equation (\ref{mk01}) implies that both positive and negative
energies are possible as taking the square root we have
        \be
        E=\pm \sqrt{p^2c^2+m^2c^4}
        \label{mk02}
        \ee

This sign ambiguity makes itself felt immediately in the Dirac
equation and led him to propose that vacuum was actually a state
in which all negative energy levels were filled by particles
obeying Fermi statistics and thus unable to accept another
particle of this type. From this point of view relativistic
quantum mechanics turned to a field theoretical description and a
procedure similar to that of non-relativistic quantum mechanics
for relativistic many body problems was essentially abandoned.

There are many different formulations of the relativistic
many-body problem \cite{7}. Conventionally they can be subdivided
to three classes. Namely, we can indicate  manifestly covariant
approaches based on the Bethe-Salpeter equation and its
generalizations, theories with direct interaction and  the mass
shell constraints approach.

The Bethe-Salpeter equation \cite{bethe}, \cite{gellman} presents
powerful and elegant tools for construction of various two body
theories which are transparently relativistic invariant. However,
a generalization of this equation to n-body case seems to be too
complicated if at all possible as far as practical applications
are concerned., see, e.g., reference \cite{5} for the case $n=3$.

A global receipt to overcome difficulties with multi time
variables was proposed by Dirac \cite{dirac1} who proposed to use
one time formulation for n-body models but to ask for existence of
realization of the Poincaré algebra on the set of solutions of the
equations of motion. The price paid for the absence of extra time
variables was the absence of manifestly  relativistic invariance.

The mass shell constraints approach \cite{droz1} shares with the
Bethe-Salpeter equation manifestly relativistic invariance. This
approach is much more easy to handle than ones based on the
Bethe-Salpeter equation. However, the problem of separation of
extra time variables is by no means trivial and was effectively
solved only for two- and three-body problems  \cite{droz2}.

In this paper we want to return to the possibility of discussing
relativistic many body problems in a framework similar to that of
the non-relativistic quantum mechanics. For this purpose we have
to deal with two problems.

\begin{itemize}

\item[a)] How can we formulate a wave equation explicitly
invariant under the Poincar\'e group but which, in an appropriate
frame of reference, involves only one time.

\item[b)] Once objective (a) is achieved how can we separate the
positive and negative parts of its solution as only the first one
will be of physical interest.

\end{itemize}

In Section 2 we deal with the first problem through a method we
developed in previous publications and in section 3 with the
second one employing a generalization of the Foldy-Wouthuysen (FW)
transformation.


\section{A Poincar\'e invariant n-body wave equation which, in a
particular frame of reference, involves only one time}

We mentioned in the introduction how we can formulate the wave
equation for a n-body non-relativistic problem starting from the
corresponding one body expression for the free particle. In the
relativistic case we have also the wave equation of
Dirac \cite{dirac2} for a single free particle given by
                \be
    \left(-i \frac{\partial}{\partial t} + \bfalpha \cdot \v p + \beta
    m\right) \psi = 0
        \label{mk03}
        \ee
where we use the usual relativistic units $\hbar=c=1$, with $\v p$
being the momentum three vector, $m$ the mass of the particle and
the matrices $\bfalpha,\beta$ are defined as in reference
\cite{dirac2}.

If we have $n$ particles of the same mass $m$, we add the index
$s=1,2\dots n$ to all the variables and an obvious Poincar\'e
invariant n-body equation can be written in the form
                \be
    \sum^n_{s=0} (\gamma^\mu_s p_{\mu s} + m) \psi =0
        \label{mk04}
        \ee
where repeated index $\mu$ are summed over the values
$\mu=0,1,2,3$ with $p_{0s}=-i \partial/\partial t_s$, $p_{js}=-i
\partial/\partial x_{js}, j=1,2,3$ and the $\gamma^\mu_s$ are
matrices related to $\bfalpha_s$ and $\beta_s$ by \cite{dirac2}
                \be
    \gamma^0_s = \beta_s, \; \gamma^i_s = \beta_s \alpha_{is}, \;
    s=1,\dots n, \; i=1,2,3
        \label{mk05}
        \ee

The $\gamma^\mu_s, p_{\mu s}$ are respectively contravariant and
covariant expressions \cite{dirac2} so that $\gamma^\mu_s p_{\mu
s}$ is a Poincar\'e scalar and thus Eq. (\ref{mk04}) is certainly
a Poincar\'e invariant $n$ particle wave equation, but it is not
satisfactory because introduces $n$ times through $p_{0s}=-i
\partial/\partial t_s, s=1,2, \dots, n$.

How can we find a formulation of many body problem, still
invariant under the Poincar\'e group but, in an appropriate system
of reference, involving only one time?

We start by denoting by $u_\mu$ unit time like four vector which
implies that there is a reference frame in which it takes the form
                \be
    (u_\mu) = (1, 0, 0, 0)
        \label{mk06}
        \ee

With the help of the four vector (\ref{mk06}) we can define the
Lorentz scalars
                \be
    \Gamma = \prod^n_{r=1} (\gamma^\mu_r u_\mu), \quad \Gamma_s =
    (\gamma^\mu_s u_\mu)^{-1} \Gamma
                        \label{mk07}
        \ee
where $(\gamma^\mu_s u_\mu)^{-1}$ eliminates the corresponding
term in $\Gamma$ so that $\Gamma_s$ is still in product form.

Instead of Eq. (\ref{mk04}) we propose now the following Lorentz
invariant one \cite{mosh1}
                \be
    \sum^n_{s=1} \Gamma_s (\gamma^\mu_s p_{\mu s} + m) \psi = 0.
        \label{mk08}
        \ee

We now introduce the total energy-momentum four vector
                \be
    P_\mu = \sum^n_{s=1} p_{\mu s}
        \label{mk09}
        \ee
and with its help define our four vector $u_\mu$ as
                \be
    u_\mu = P_\mu (-P_\tau P^\tau)^{\ahalf}
       . \label{mk10}
        \ee

We immediately see that when the center of mass of our n-body
system is at rest, \ie, $P_i=0, i=1,2,3$ our four vector $u_\mu$
takes the form Eq. (\ref{mk06}) in which the wave equation
(\ref{mk08}) becomes
                \be
    \bigg[\Gamma^0 \sum^n_{s=1} p_{0s} + \sum^n_{s=1} \Gamma^0_s
    (\bfgamma_s \cdot \v p_s)\bigg] \psi = 0
        \label{mk11}
        \ee
where bold face letters mean three dimensional vectors and
                \be
    \Gamma^0 = \prod^n_{r=1} \gamma^0_r, \quad \Gamma^0_s =
    (\gamma^0_s)^{-1} \Gamma^0.
        \label{mk12}
        \ee

Multiplying Eq. (\ref{mk11}) by $\Gamma^0$ and making use of Eqs.
(\ref{mk05}), (\ref{mk09}), (\ref{mk12}) we obtain
                \be
    [-P^0 + \sum^n_{s=1} (\bfalpha_s \cdot p_s + m \beta_s)]\psi =
    0
        \label{mk13}
        \ee
where we used a metric in which $P_0=-P^0$ and the latter is the
zero component of the four vector $P^\mu$, \ie, the total energy of
the system.

So far we have obtained a Poincar\'e invariant wave equation for a
system of non-interacting particles which in the frame of
reference in which $\v P=0$ takes the form (\ref{mk13}).

We wish now to consider interactions and for simplicity we will
consider them to depend only on the relative coordinates
                \be
    x^{st}_\mu \equiv x_{\mu s} - x_{\mu t}, \; \mu =0,1,2,3
        \label{mk14}
        \ee

We want that these relative coordinates become purely spatial ones
in the frame of reference where the total momentum $\v P=0$ and
this is easily achieved through the use of the $u_\mu$ four vector
of (\ref{mk06}) by defining
                \be
    x^{st}_{\perp \mu} \equiv x^{st}_\mu - (x^{st}_\tau u^\tau)
    u_\mu
        \label{mk15}
        \ee
because when $u_\mu$ takes the form (\ref{mk06}) the
$x^{st}_\mu$ becomes
                \be
    x^{st}_0 = 0 \;,\; x^{st}_i = x_{is} - x_{it}, \quad i=1,2,3
        \label{mk16}
        \ee

As we want our potential to be Poincar\'e invariant it is
sufficient to make it a function of
                $
    x^{st}_\mu x^{s't'\mu}
        $
where repeated index $\mu$ are summed over their values
$\mu=0,1,2,3$. Thus we restrict ourselves to potentials of the
form
              \be
   V= V (x^{st}_\mu x^{s't'\mu})
        \label{mk18}
        \ee
which in the frame of reference where $\v P =0$ becomes $V(\v
x^{st}\cdot \v x^{s't'})$ with the bold face letters indicating
spatial relative vectors, \ie
                \be
    \v x^{st} = \v x_s - \v x_t
        \label{mk19}
        \ee

We are now in a position to write of Poincar\'e invariant wave
equation for as of $n$ particle of the same mass and of spin
$\ahalf$ as
                \be
                \bigg[ \sum^n_{s-1} \Gamma_s \cdot  (\gamma^\mu_s p_{\mu
                s} + m) + \Gamma V (x^{st}_\mu x^{s't'\mu})\bigg]
                \psi =0
        \label{mk20}
        \ee
which in the frame of reference in which the center of mass of the
system is at rest, \ie, $\v P=0$ leads to the Hamiltonian equation
                \be
    H\psi  \equiv \bigg[ \sum^n_{s=1} (\bfalpha_s \cdot \v p_s + m
    \beta_s) + V(\v x^{st} \cdot \v x^{s't'})\bigg] \psi = E\psi
        \label{mk21}
        \ee
where we replaced $P^0$ by the total energy $E$. So we can denote
the square bracket in (\ref{mk21}) as our Hamiltonian.

Equation (\ref{mk21}) is not the end of our story because even for
the one particle case it involves both positive and negative
energies and it is the former ones that will be of interest to us.
Fortunately Foldy and Wouthuysen \cite{FW}  gave us a procedure to
separate the positive and negative energy parts for the one
particle case and in the next section we proceed to generalize it
for the $n$ particle system.

\section{The FW transformation for the many
body Hamiltonian}

To be able to derive the FW transformed Hamiltonian for the $n$
body system we need to review briefly the corresponding analysis
for the one body case.

The Hamiltonian is then \cite{bjorken}
                \be
                H \equiv {\mathcal{O} + \mathcal{E}} + V
        \label{mk22}
        \ee
where the odd part $\mathcal{O}$ is $\bfalpha \cdot\v p$, the even one
${\mathcal{E}}=\beta m$ and $V$ is the potential being a scalar function
of the one variable $\v x$.

We now follow FW \cite{FW} and, in particular, the book of Bjorken
and Drell \cite{bjorken} which states that it is possible to find
a unitary operator

                \be
                U = \exp(iS)
                \label{mk23}
        \ee
that allows us to transform $H$ in a series of inverse powers of
the mass $m$ corresponding to the positive energy part of $H$.

The $S$ in (\ref{mk23}) is given as

                \be
    S = -\frac{i\beta}{2m} (\ocal + \ocal'+\ocal'')
        \label{mk24}
        \ee
with
                \be
            {\ocal} = \bfalpha\cdot \v p, \quad {\ocal}' = \beta
             \frac{[\bfalpha\cdot \v p, V]}{2m} , \; {\ocal}''=-
            \frac{(\bfalpha \cdot \v p) p^2}{3m^2}
        \label{mk25}
        \ee
and the new positive energy Hamiltonian $H'$ is now given as
                \be\begin{array}{l}
                H'=\hat H + V, \; \hat H = \beta \bigg(m+\frac{p^2}{2m}
                - \frac{p^4}{8m^3}\bigg)\\ + \frac{1}{4m^2} \v s
                \cdot \bigg[ (\v p \times \v E) - (\v E\times \v
                p)\bigg] + \frac{1}{8m^2} \nabla^2 V
                        \end{array}\label{mk26}
        \ee
where one makes the expansion only through order (kinetic
energy$^2$/m$^3$) and [(energy)(field energy)]/m$^2$. The field
strength $\v E$ is given by
                \be
                \v E =- \nabla V
       \label{mk27}
        \ee
and $\v s$ is the spin of the particle $\v s=- i/{4} {\bfalpha}\times{\bfalpha}$.

 Once we have the above well known results we pass to obtaining the
corresponding ones for the n-body problem

We start with the two body case where
                        \be
        H = H_1 + H_2 + V (\v x_1, \v x_2)
        \label{mk28}
        \ee
with
                            \be
    H_s = \bfalpha_s \cdot \v p_s + \beta_s m, \; s=1,2
        \label{mk29}
        \ee

Rewriting $H$ as
                                \be
        H = [H_1 + V (\v x_1 , \v x_2)] + H_2
        \label{mk30}
        \ee
we can first apply to it the unitary transformation
                            \be
        U_1 = \exp (i S_1)
        \label{mk31}
        \ee
where $S_1$ is given by $S$ of (\ref{mk24}) where $\beta,
\bfalpha, \v p$ have the index 1 \ie $\beta_1, \bfalpha_1,  \v
p_1$ to indicate its dependence on particle 1. As $S_1$ only
depends on the variables of particle 1 it is clear that it does
not affect $H_2$ in (\ref{mk30}) and for $(H_1 +V)$ it gives the
result of (\ref{mk26}) with an index 1 for the variable \ie
               \be
               \exp (i S_1) H \exp (-iS_1)  = \hat H_1 + (H_2 + V)
        \label{mk32}
        \ee
where $\hat H_1$ is given by (\ref{mk26}) with index one for all
the variables.

We now apply the unitary transformation of the second particle \ie
                \be
                U_2 = \exp (iS_2)
        \label{mk33}
        \ee
where $S_2$ is given by $S$ of  (\ref{mk24}) where $\beta, \v x,
\v p$ have the index 2 \ie $\beta_2, \bfalpha_2, \v p_2$. It is
clear that the unitary transformation $U_2$ has no effect on $\hat
H_1$ as the only terms it could affect are $(\nabla^2_1 V/8m^2)$
and $\frac{1}{4m^2} \v S
                \cdot \bigg[ (\v p \times \v E) - (\v E\times \v
                p)\bigg]$ and this would be of higher order that the ones we accept.
We have then that $U_2$ only acts on $(H_2+V)$ giving us

                \be
        U_2 (H_2 + V) U^\dag_2 = \hat H_2 + V
        \label{mk34}
        \ee
where operator $\hat H_2$ is the one given in (\ref{mk26}) where
all the variables have index 2. Thus if consider $U_2U_1$ as our
unitary transformation we have
                \be
               U_2 U_1 (H_1 + H_2 + V) U_1^\dag U^\dag_2 = \hat
               H_1 + \hat H_2 + V
                        \label{mk35}
        \ee

The procedure for the two particle problem immediately suggests
that for the n-body case where $V(\v x_1, \v x_2, \dots \v x_n)$
we can carry out the transformation

                \be
           H' = U_n U_{n-1} \cdots U_2 U_1 H U^\dag_1 U^\dag_2 \cdots
           U^\dag_n = \hat H_1 + \hat H_2 +\cdots+ \hat H_n + V
        \label{mk36}
        \ee
where
                       \be\begin{array}{l}
        H_t = \beta_t \bigg(m+ \frac{p^2_t}{2m} -
        \frac{p^4_t}{8m^3}\bigg) + \frac{1}{4m^2} \v s_t \cdot (\v p_t \times \v E_t-\v
        E_t \times \v p_t )\\ + \frac{1}{8m^2}
        \nabla^2_t V,\ \ \ \ \ t=1,2,\cdots n\end{array}
        \label{mk37}
        \ee

We must still keep in mind that in our equation (\ref{mk34}) we
have to take into account that the total momentum $\v P=0$, which
we can achieve by passing from our coordinate system to the
Hamilton-Jacobi one as will be indicated in the examples discussed
in the following section.

Thus we extend the approximate FW transformation to the case of multi
particle relativistic wave equations. For exact FW transformations for one- and two-particle
systems see refs. \cite{viola}.

\section{The two-body problem}

In the case of two particles when the total momentum $\v P=0$ we
have that
\be \v p_1+\v p_2=0,\ \ {\mbox {or}}\ \ \v p_1=-\v p_2\v\equiv \v
p\label{mk38}\ee
and we shall denote the corresponding relative coordinate vector
as $\v r=\v r_1-\v r_2$. Besides, for simplicity, we shall take
the potential $V $ as of harmonic oscillator one, \ie,
$V=\frac{m\omega^2r^2}{4}$.

From (\ref{mk37}) and using the notation $\v p, \v r$ introduced
in the above we obtain that
\be\begin{array}{l}  H' = (\beta_1+\beta_2) \bigg(m+\frac{p^2}{2m}
                - \frac{p^4}{8m^3}\bigg) +V\\+ \frac{1}{4m^2}\bigg( \v
                s_1+\v s_2\bigg)
                \cdot \bigg[ (\v p \times \v E) - (\v E\times \v
                p)\bigg] + \frac{1}{4m^2} \nabla^2 V\end{array}\label{mk40}\ee
as
\be\v E_1=\v E_2=\v E=-\nabla V=-\frac{m\omega^2\v r}{2}\ \mbox {and}\
\nabla^2_1 V=\nabla^2_2 V=3\frac{m\omega^2}{2}\label{mk41}\ee
when $V=\frac{m\omega^2\v r^2}{4}$. Introducing then the total
orbital angular momentum
 $\v L=\v r\times \v p$ and using (\ref{mk41})
we reduce $H'$ to the form
\be  H' = (\beta_1+\beta_2) \bigg(m+\frac{p^2}{2m}
                - \frac{p^4}{8m^3}\bigg) +\frac{m\omega^2r^2}{4}+ \frac{\omega^2}{4m} \v
                S\cdot \v L + 3\frac{\omega^2}{8m}\label{mk43}\ee
where
 \be \v S =\v s_1+\v s_2\label{mk44}\ee
 is the total spin vector and so the spin values can only be $1$
 or $0$.

 We note also that in Eq. (\ref{mk43}) besides terms familiar in
 non-relativistic quantum mechanics there are the matrices
 $\beta_1$ and $\beta_2$. They can be chosen in the form of the
 direct products \cite{mosh1}
 \be
 \beta_1=\left(\begin{array}{cc}I_2&0\\0&-I_2\end{array}\right)
 \otimes\left(\begin{array}{cc}I_2&0\\0&I_2\end{array}\right),\
\beta_2=\left(\begin{array}{cc}I_2&0\\0&I_2\end{array}\right)
 \otimes\left(\begin{array}{cc}I_2&0\\0&-I_2\end{array}\right)\label{mk45}\ee
 where $I_2$ is the $2\times 2$ unit matrix, and so
 \be\beta_1+\beta_2=2\left[\begin{array}{ccc}I_4&&\\&0_8&\\&&-I_4\end{array}\right]\label{mk46}\ee
 when $I_4$ and $0_8$ are the $4\times 4$ unit matrix and $8\times
 8$ zero matrix correspondingly.

 Our only interest in the positive energy part of the wave
 function which implies that $\beta_1+\beta_2$ should be replaces
 by 2 and thus finally we have to deal with the expression
\be  H' = \bigg(2m+ 3\frac{\omega^2}{8m}\bigg)+\bigg(\frac{p^2}{m}
                 +\frac{m\omega^2r^2}{4}+ \frac{\omega^2}{4m} \v
                S\cdot \v L\bigg)- \frac{p^4}{4m^3}\label{mk47}\ee

As the second parenthesis correspond to an harmonic oscillator
with spin-orbit coupling whose  eigenfunctions and eigenvalues are
very well known we can use the former as a complete basis to
convert $H'$ into a numerical matrix.

Before proceeding to give a procedure to calculate the eigenvalues
for $H'$ of (\ref{mk47}) it is convenient to make the canonical
transformation
\be {\bfpi}={\sqrt\frac{2}{m\omega}}\v p,\ \v {\bfrho}=\sqrt
{\frac{m\omega}{2}}\v r \label{mk471}\ee to write $H'$ as
 \be H' =2m+
3\frac{\omega^2}{8m}+\frac{\omega}{2}\bigg(\pi^2+\rho^2\bigg)+\frac{\omega^2}{4m}
\v
                S\cdot \v L- \frac{\omega^2}{16m}{\pi ^4}\label{mk472}\ee
where in the second parenthesis we have the oscillator hamiltonian
of unit frequency and the spin-orbit term remains uncharged as
$\v r \times\v p=\v {\bfrho}\times\v {\bfpi}$ while $p^4$ is replaced
by $\frac14m^2\omega^2\pi^4$.

To convert $H'$ into a numerical matrix we can use the states of
the harmonic oscillator and states of squared total angular
momentum $J^2$, total orbital momentum $L^2$ and total spin $S^2$,
\i.e.,
\be|n l,\bigg(\frac12\frac12\bigg) S; j, m>=\sum_{\mu,\sigma}<l
\mu, S \sigma|j m>|n l \mu>|\bigg(\frac12\frac12\bigg) S
\sigma>\label{mk473}\ee where $J^2,\ L^2$ and $S^2$ commute with
the Hamiltonian (\ref{mk472}) and so are integrals of motion. The
kets $|n l \mu>$ are those of the harmonic oscillator of unit
frequency \cite{mosh2} and $ |\bigg(\frac12\frac12\bigg) S
\sigma>$ are those of the total spin.

The numerical matrix we want to determine has then the elements
\be\begin{array}{c}<n' l,\bigg(\frac12\frac12\bigg) S; j, m|H'|n
l,\bigg(\frac12\frac12\bigg) S, j,
m>\\=\bigg(2m+\frac{3\omega^2}{8m}+\omega\bigg(2n+l+\frac32\bigg)+
\frac{\omega^2}{8m}[j(j+1)-l(l+1)-s(s+1)]\bigg)\delta_{n
n'}\\
-\frac{\omega^2}{16m}<n' l'|\pi^4|n l>\end{array}\label{mk474}\ee

The last term in (\ref{mk474}) can then be defined starting with
the relation $<n'l|\pi^4|n l>=<n' l|\pi^2|n'' l><n'' l|\pi^2|n'
l>$  and using  the expression for $<n' l|\pi^2|n l>$ given in p.
7, Eq. (3.11) of reference \cite{mosh2}. Thus we get that
\be\begin{array}{c}<n'l|\pi^4|n
l>=\sqrt{n(n-1)(n+l+\frac12)(n+l-\frac12)} \delta_{n' \
n-2}\\+(4n+2l+1)\sqrt{n(n+l+\frac12)}\delta_{n'\
n-1}\\+[(2n+l+\frac32)(2n+l+\frac52)+2n(n+l+\frac12)]\delta_{n'\
n}\\+(4n+2l+5)\sqrt{(n+1)(n+l+\frac32)}\delta_{n'\
n+1}\\+\sqrt{(n+1)(n+2)(n+l+\frac32)(n+l+\frac52)}\delta_{n'\
n+2}\end{array}\label{mk475}\ee

We note that as we are using units in which $\hbar=c=1$, the
dimensionless term $\omega/m$ becomes in the c.g.s. units
\be\frac{\omega}{m}\to\frac{h\omega}{mc^2}\label{mk476}\ee and can be treated as a
small parameter. In following section we compare the results of our
analysis with the experimental spectrum of the bottonium and charmonium masses.

\section{
Energy spectrum of the two body problem}

We start with  the eigenvalue problem \be
H'\psi=E_{nl}\psi\label{mk001}\ee for the two-body hamiltonian
(\ref{mk474}).

 To find the related energy spectrum we are
 supposed to diagonalize matrix (\ref{mk475})  which can be
 done using numerical methods. Moreover, for sufficiently
 small coupling constant  (\ref{mk476}) it is possible to apply the
 standard perturbation theory and express the eigenvalues of $H'$ (\ref{mk474}) in power series of
 $\nu=\omega/m$:
 \be E_{nl}-2m-\frac{3\omega^2}{16m}=E^0_{nl}+E^1_{nl}+E^2_{nl}+\cdots\label{mk009}\ee
where the non-perturbed levels $E^0_{nl}$ are linear in $\nu$,
 \be E^0_{nl}=m\nu\left(2n+l+\frac32\right)\label{mk008}\ee
 while $E^1_{nl}$ and $E^2_{nl}$ are quadratic and cubic in $\nu$ respectively. Moreover,
 the first and second perturbations of energy spectrum can be expressed via the elements of
 the perturbing matrix $<n'l'j'|K|jnl>=\frac{\omega^2}{8m}[j(j+1)-l(l+1)-s(s+1)]\delta_{n
n'}\delta_{l
l'}
-\frac{\omega^2}{16m}<n' l'|\pi^4|n l>\delta_{j
j'}$ as follows \cite{2}
\be E^1_{nlj}=<nlj|K|nlj>, \  E^2_{nlj}=\sum_{n'\neq n,l'\neq l}\frac{<n'l'j |K|n'l'j>^2}{E^0_{nl}-E^0_{n'l'}}\ee
Then, using (\ref{mk475}), (\ref{mk009}) and (\ref{mk008}) we obtain
  \be\begin{array}{l}
 E_{nl}=2m+\frac{3\omega^2}{16m}+\omega
 (2n+l+\frac32)+
\frac{\omega^2}{8m}[j(j+1)-l(l+1)-s(s+1)]\\-\frac{\omega^2}{4m}[(2n+l+\frac32)(2n+l+\frac52)
 +2n(n+l+\frac12)]\\-\frac{\omega^3}{16m^2}\bigg(2n+l+\frac32\bigg)
 \bigg(9n\bigg(n+l+\frac12\bigg)+
 2\bigg(2n+l+\frac52\bigg)\bigg(2n+l+\frac{11}{4}\bigg)\bigg)\end{array}\label{mk002}\ee

 Formula (\ref{mk002}) describes the spectrum of relativistic two
 body systems  with the harmonic oscillator potential. We compare it with the energy spectrum
 of two quark systems (mesons) and find that  it presents a rather realistic qualitative
 distribution of bottonium and charmonium masses. For example, setting $m=4.7\ GEV$
 (i.e., supposing $m$ be equal to the bottonium quark mass) and choosing the
 dimensionless coupling constant
 $\nu=0,19$ we obtain from (\ref{mk002}) the following values for the mass differences
 of the bottonium states: $\xi_{b0}-\Upsilon(1S)=0.411 \ GEV, \xi_{b1}(1P)-
 \Upsilon(1S)=0.423\  GEV,
 \xi_{b2}(1P)-\Upsilon(1S)=0.443 \ GEV$ while the experimental data are
 \cite{data} 0,400, 0,432 and 0,453 $GEV$
 respectively.

 To obtain a better agreement with experimental data we consider anharmonic oscillator potential
\be V=\frac{m\omega^2r^2}{4}+V', \ \ V'=-\frac{\alpha m\omega^4r^4}{64}\label{mk003}\ee
where $\alpha$ is a dimensionless interaction constant. In addition,
we take into account relativistic corrections up to order $\nu^3$ for the approximate
Hamiltonian $H'$ which needs continuation of the FW reduction. We omit the routine
calculations which are analogous to ones given in Section 3
 and present the resulting transformed Hamiltonian using variables (\ref{mk471}):
\be\begin{array}{l}H'=2m+\frac{3\omega^2}{8m}+\frac{\omega}{2}\bigg(\pi^2+
\rho^2\bigg)+\frac{\omega^2}{4m}
\v S\cdot \v L- \frac{\omega^2}{16m}(\pi ^4+\alpha\rho^4)\\
+\frac{\omega^3}{32m^2}\left(\frac{\pi^6}{2}+(2-5\alpha)\rho^2-
(3\pi^2+2\rho^2)\v S\cdot \v L-(\v S\cdot \bfpi)^2+(\v S^2-8)\pi^2\right)

\end{array}
\label{mk004}\ee

We need only diagonal matrix elements for terms of order $\omega^3/m^2$, placed at the second line of equation (\ref{mk004}).
 They can be easily found starting with matrix representation
for $\rho^2$ and $\pi^2$ given in page 7, Eq. (3.11) of reference \cite{mosh2} and
the representation for $\v S\cdot \bfpi$ in the spherical spinor basis given in
pages 422-423 of reference \cite{FN} :
\[\begin{array}{c}<nl|\rho^2|nl>=<nl|\pi^2|nl>=2n+l+\frac32,\\
<nl|\rho^4|n
l>=(2n+l+\frac32)(2n+l+\frac52)+2n(n+l+\frac12)],
\\
<nl|\pi^6|nl>=4(n+l+\frac12)(2n+l+1)\\+
(2n+l+\frac52)\left(2(n+1)(n+l+\frac32)+(2n+l+\frac32)^2\right),\\
<nlj|(\v S\cdot\bfpi)^2|nlj>=\left(1-\frac12(j-l)^2+\frac{j-l}{2(2j+1)}\right)
<nl|\pi^2|nl>
\end{array}\]
and the related energy values $E_{nlj}$ are described by the following formula
 \be\begin{array}{l} E_{nlj}= 2m+\frac{(3-\v S^2)\omega^2}{8m}+\omega
 (2n+l+\frac32)+\frac{\omega^2}{8m}\left[j(j+1)-l(l+1)\right]\\-\frac{(1+\alpha)\omega^2}{16m}\left[\left(2n+l+\frac32\right)\left(2n+l+\frac52\right)+
 2n\left(n+l+\frac12\right)\right]\\
  -\frac{\omega^3}{256m^2}\left(2n+l+\frac32\right)
 \left[(1-\alpha)^2\left(2\left(2n+l+\frac52\right)^2+4n\left(2n+l+\frac12\right)\right)\right.\\
 \left.+(1+\alpha)^2\left((n+1)\left(n+l+\frac32\right)-
 \frac12\left(2n+l+\frac12\right)\right)\right]\\-
 \frac{\alpha\omega^3}{32m^2}\left(2n+l+\frac32\right)(j(j+1)-l(l+1)+5-{\bf S}^2)\\+
 \frac{\omega^3}{64m^2}\left(4n(n+l+\frac12)(2n+l+1)+
(2n+l+\frac52)\left(2(n+1)(n+l+\frac32)\right.\right.\\\left.\left.
+(2n+l+\frac32)^2\right)\right)\\
+\frac{\omega^3}{64m^2}(2n+l+\frac32)\left[3l(l+1)-3j(j+1)+(j-l)^2+\frac{l-j}{2j+1}+5{\bf S}^2-14\right].
\end{array}\label{mk006}\ee

 We compare the spectrum (\ref{mk006}) with the experimental bottonium and charmonium mass
 spectra \cite{data}. We set $m=4.7 GEV$ for the bottonium  and $m=1.4$ for charmonium cases and
 use MAPLE software to find the coupling constants $\omega$ and $\alpha$ which correspond to
 the minimal deviation of the spectrum (\ref{mk006}) from experimental data.
 The results of our investigations are given in the following tables.

 \vspace{1cm}

 \begin{center} Table 1. Experimental spectra of the bottonium and
 our model results (in GEV), $m=4.7, \omega=0.378,\  \alpha=4.7$
 \end{center}
 \begin{tabular}{lll}
 &Experimental&Theoretical\\
$l=0,j=1$&&\\
$\ \ \Upsilon(2S) -\Upsilon(1S)$&0.563&0.574\\
$\ \ \Upsilon(3S)-\Upsilon(1S)$&0.895&0.963\\
$\ \ \Upsilon(4S)-\Upsilon(1S)$&1.119&1.142\\
$l=1,j=0$&&\\

                        $\ \ \xi_{b0}(1P)-\Upsilon(1S)$&0.400&0.310\\
$\ \ \xi_{b0}(2P)-\Upsilon(1S)$&0.772&0.797\\
$l=1,j=1$&&\\
$\ \ \xi_{b1}(1P)-\Upsilon(1S)$&0.432&0.316\\

                       $\ \
\xi_{b1}(2P)-\Upsilon(1S)$&0.794&0.801\\
$l=1,j=2 $&&\\
$\ \ \xi_{b2}(1P)-\Upsilon(1S)$&0.453&0.326\\
$\ \ \xi_{b1}(2P)-\Upsilon(1S)$&0.808&0.808
\end{tabular}.

\newpage

%
\begin{center} Table 2. Experimental spectra of the charmonium and
 our model results (in GEV), $\omega=0.45, \alpha=1.2, m=1.4$
 \end{center}
 \begin{tabular}{lll}
 &Experimental&Theoretical\\
$l=0,j=1, s=0$&&\\
$\ \ \eta'_c(2S) -\eta_c(1S)$&0.603&0.604\\
$l=0,j=1, s=1$&&\\
$\ \ \Psi(2S)-J/\Psi(1S)$&0.589&0.589\\
$\ \ \Psi(3S)-J/\Psi(1S)$&0.943&0.925\\
$l=1,j=0,s=1$&&\\
$\ \ \xi_{c0}(1P)-J/\Psi(1S)$&0.308&0.372\\
$l=1,j=1, s=1$&&\\
$\ \ \xi_{c1}(2P)-J/\Psi(1S)$&0.413&0.390\\
$l=1,j=2, s=1 $&&\\
$\ \ \xi_{c2}(3P)-J/\Psi(1S)$&0.459&0.422
\end{tabular}.

We see that in average the mass spectrum predicted by our very simple model
is in rather good accordance with the experimental data.
It is possible to obtain a better agreement with experimental data
changing the quark masses by effective ones which are additional free parameters.

\section{Conclusion}

We propose a relativistic Hamiltonian for an n-body problem in
which all the masses are equal and all spins take value 1/2.
Discussing the problem in the frame of reference in which the
total momentum $\v P=0$, we were able to extend the FW
transformation to n-body case and separate the positive energy
part of the Hamiltonian.  Examples of two body systems
are discussed in more detail.

The proposed approach admits a straight forward generalization to
the case of particles with different masses and spins and is valid
for more general form of the interaction potential. In the present
paper we discuss only the main ideas and demonstrate it
effectiveness using the simplest interaction model. Nevertheless,
even this very straight forward model predicts a rather realistic
bottonium and charmonium spectra presented in the Appendix. It looks rather
curiously that starting with a an interaction potential which is
not well grounded physically and using  only two free parameters
$\omega/m$ and $\alpha$ it is possible to obtain a good qualitative and also
relatively good quantitative description of bottonium and charmonium masses.

\end{document}